\documentclass[10pt,a4paper]{article}
\usepackage{natbib}
\usepackage{graphicx}
\usepackage{xcolor}

\usepackage{authblk}

\usepackage{amsmath,amssymb}

\title{Volatility Forecasting with 1-dimensional CNNs via transfer learning}
\author[1]{Bernadett Aradi\thanks{aradi.bernadett@inf.unideb.hu}}
\author[2]{G\'abor Petneh\'azi\thanks{gabor.petnehazi@science.unideb.hu}}
\author[1]{J\'ozsef G\'all\thanks{gall.jozsef@inf.unideb.hu}}
\affil[1]{Department of Applied Mathematics and Probability Theory, University of Debrecen}
\affil[2]{Doctoral School of Mathematical and Computational Sciences, University of Debrecen}
\date{}

\begin{document}
\maketitle

\begin{abstract}
Volatility is a natural risk measure in finance as it quantifies the variation of stock prices. A frequently considered problem in mathematical finance is to forecast different estimates of volatility. What makes it promising to use deep learning methods for the prediction of volatility is the fact, that stock price returns satisfy some common properties, referred to as `stylized facts'. Also, the amount of data used can be high, favoring the application of neural networks. We used 10 years of daily prices for hundreds of frequently traded stocks, and compared different CNN architectures: some networks use only the considered stock, but we tried out a construction which, for training, uses much more series, but not the considered stocks. Essentially, this is an application of transfer learning, and its performance turns out to be much better in terms of prediction error. We also compare our dilated causal CNNs to the classical ARIMA method using an automatic model selection procedure.
\end{abstract}


\section{Introduction}
Volatility (the variation of prices) has great importance in finance as a natural risk measure. However, it is not observable, so the notion covers different estimates of the true price variability.\\
Volatility estimates might be computed from high frequency intraday data \citep{andersen2003modeling}, or from the daily range of prices \citep{chou2010range}, but they are probably most often computed simply as the standard deviation of daily returns. There is also a considerable confusion around the concept of volatility \citep{taleb2007we}. We chose to examine a pretty standard estimate of stock volatility: the 21-day moving standard deviations of daily logarithmic returns.\\
In this study, we aim to build a deep learning based general forecasting model to predict future values of these volatility estimates. More specifically, we train a one dimensional convolutional neural network on multiple stocks’ volatility history, and compare its forecasting performance to that of models that were trained on a single stock’s data only. We show that a general model might perform comparably or even better, than the stock specific models, which may justify the application of deep learning methods to financial forecasting.

\section{Generality of stock price volatility}
Stock market returns are often modeled using random walk hypothesis, even though it was invalidated empirically by more studies, see, e.g.\ \citet{lo1988stock}. However, if we intend to use deep learning methods for forecasting volatility, it is more important, that these returns have some documented common properties, usually referred to as \emph{stylized facts} [\citet{cont2001empirical}, \citet{engle2001good}]. 
These patterns in the behaviour of the different assets' returns can come very handy, and they suggest that the variability of asset prices is forecastable. \\
Some of these similarities are the following. Volatilities cluster (persistence), that is, they display positive autocorrelation. Returns and volatilities are negatively correlated, so financial asset returns are typically more volatile during recessions. Also, positive and negative shocks in returns have different impact on volatility. In general, trading volume is positively correlated with volatility. Volatility exhibits mean reversion as well, meaning that, in the long run, it corverges to a normal volatility level. Finally, exogenous variables (e.g., other assets or deterministic events) can have an impact on volatility. Thus, there might be general economic factors that influence the volatilities of individual stocks.\\
These patterns, in accordance with \citet{schwert1989does}, stating that there is weak evidence that macroeconomic volatility can help predict stock volatility, imply that the price development of different stocks might have common driving forces. Furthermore, knowledge extracted from the price behaviour of some assets might be useful to describe that of some other assets. \\
This key idea has already been applied by \citet{sirignano2019universal}, however, to forecast only the direction of price movements, using a high frequency database of market quotes and transactions. They found that a universal model trained for all stocks outperforms asset-specific models. The authors claim it is evidence of a universal price formation mechanism. Those previous findings encourage us to study if the remarkable generality in stock volatility formation can help volatility forecasting. That is, we study if the volatility history of multiple stocks can be used in a joint system to predict the future volatility of the individual securities.

\section{Convolutional neural networks}
Convolutional neural networks are most often used with images. \citet{lecun1989backpropagation} applied them to handwritten digit recognition and thereby launched a revolution in image processing. Since then, better and better CNN architectures have been proposed to solve difficult image classification and object detection problems, and now their performance is often comparable to that of human experts \citep{russakovsky2015imagenet}. Due to this undeniable success in computer vision, we usually associate CNNs with images.\\
However, they can also be used with other data. We can even let go of the restriction of having two dimensions: 1 or 3 dimensional convolutions can be used in pretty much the same way.\\
Actually, the time delay neural network of \citet{waibel1989phoneme} was the first ever CNN---a convolutional network applied to the time dimension.\\
CNNs can extract local features, which is a useful property, since variables spatially or temporarily nearby are often highly correlated \citep{lecun1995convolutional}. These features are learnt by backpropagation, thus CNNs build a perfectly self-acting feature extractor. They are also very efficient in the number of parameters, due to the weight sharing in space or time.\\
So, convolutional neural networks can be applied to time series and to images in a similar manner. Extracting local patterns might be just as useful in the time domain.\\
\citet{bai2018convolutional} claim that while the general belief is that sequence modeling is best handled by recurrent neural networks, convolutional neural networks might even outperform them, considering generic architectures. In the light of this, it is not surprising, that in the past few years several CNN architectures were successfully applied to time series forecasting.\\
\citet{mittelman2015time} used fully convolutional networks to time series modeling, replacing the usual subsamplings and upsamplings by upsampling the filter of the $l^{th}$ layer by a factor of $2^{l-1}$. \citet{yi2017grouped} presented structure learning algorithms for CNN models, exploiting the covariance structure of multiple time series. \citet{binkowski2017autoregressive} proposed a CNN architecture to forecast multivariate asynchronous time series. \citet{borovykh2017conditional} applied a CNN inspired by the WaveNet \citep{oord2016wavenet}, using dilated convolutions. Dilations allow an exponential expansion of the receptive field, without loss of coverage \citep{yu2015multi}.\\
Some of the mentioned studies applied the proposed methods to financial data\-sets, since they often pose a challenge to traditional time series forecasting algorithms.\\
In this article, we are going to apply convolutional networks to series of stock return volatilities, and use the learned patterns to predict the subsequent values of the series.\\
A CNN seems a good choice for learning from multiple time series, since it can recognize different local patterns in the data. It has enough complexity to account for various time series phenomena---many more than what might be present in a single time series. We may also expect this jointly learnt model to help avoid overfitting---instead of just memorizing the given time series history, the algorithm might learn general time series behavior from a much richer source.

\section{Data}
We have downloaded 10 years (from the beginning of 2009 to the end of 2018) of daily prices for hundreds of frequently traded stocks---constituents of the S\&P 500 stock market index. The dataset was obtained through the Python module of Quandl\footnote{https://www.quandl.com/data/WIKI}. Volatilities were estimated as 21-day moving standard deviations of daily logarithmic returns. The estimates were annualized by multiplying each value by the square root of 252. After removing stocks with more than 10 missing observations, 440 volatility series remained. Each series was split to overlapping 64-day subseries, which were fed to the algorithm to predict the following, 65th value. The data was standardized by subtracting the total mean and dividing by the total standard deviation of the whole training set.

\section{Network Architecture}
Motivated by the recent successes of CNNs in time series forecasting, we chose to use a dilated causal 1 dimensional convolutional neural network. The inputs to this network are 64-step sequences of the computed volatility estimates, while the outputs are the subsequent, 65th value, so that we look one step ahead into the future. The causal convolution means that the output at one point in time only depends on inputs up to that point, and the data is padded, such that the input and the output have the same length. Dilated convolution (or convolution with holes) makes the filter larger by dilating it with zeros. The dilation rate of the $l^{th}$ layer is set to $2^{l-1}$, which allows an exponential receptive field growth, and enables a relatively shallow network to look into a relatively distant past. We use 6 causal convolutional layers with exponentially increasing dilation rates. Each layer uses 8 filters with a kernel size of 2, and a $relu$ activation function. It is then followed by a final convolutional layer with a kernel size of 1 and a single filter, so that the output shape matches that of the given time series sequence. The networks were trained for 300 epochs, using the $adadelta$ \citep{zeiler2012adadelta} optimizer.

\section{Experiments}
We have randomly chosen 10 stocks, and we used two CNN forecasters for each. The first (so-called individual) model learns from the volatility history of the given stock only. The second (joint) model learns from all stocks' volatilities, except the chosen 10 stocks. It means that the second model learns from more than 400 times as much data, however it totally disregards the time series that we are forecasting. This method can be considered as a way of using \emph{transfer learning}: in the training set we have different, but very similar time series than in the test set. \\
We also applied an ARIMA model, in order to extend the comparison to a simpler and more classical time series forecasting method. The models were trained and tested on separate time periods: the first 70\% of the available nearly 10 years long time period was used for training the models, while the remaining 30\% was the evaluation set. We produced one-day-ahead forecasts, and compared the models’ performance in terms of forecast error and directional accuracy.

\section{Results}
\begin{figure}
\centering
\includegraphics[width=\textwidth,height=\textheight,keepaspectratio]{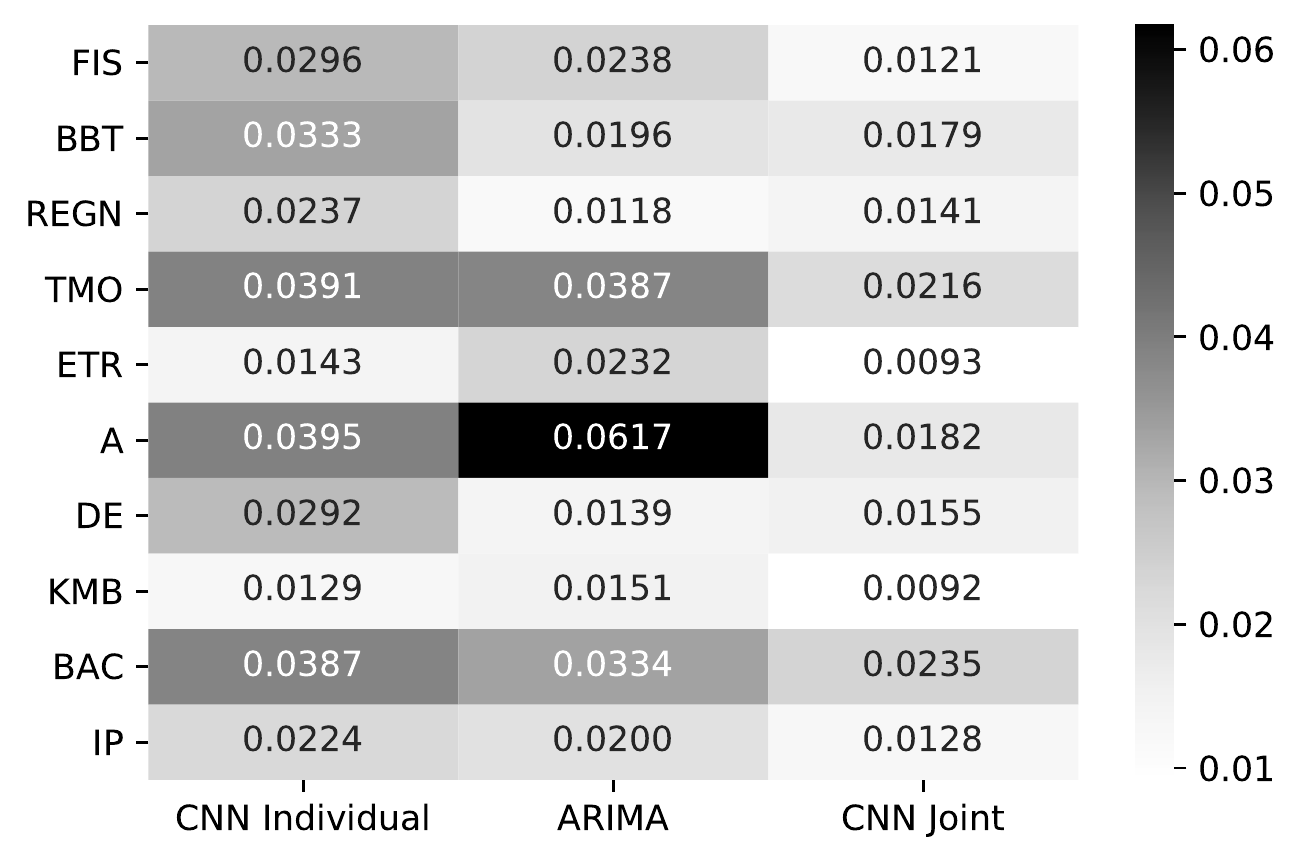}
\caption{RMSE Scores}
\label{fig:rmse_scores}
\end{figure}

\begin{figure}
\centering
\includegraphics[width=\textwidth,height=\textheight,keepaspectratio]{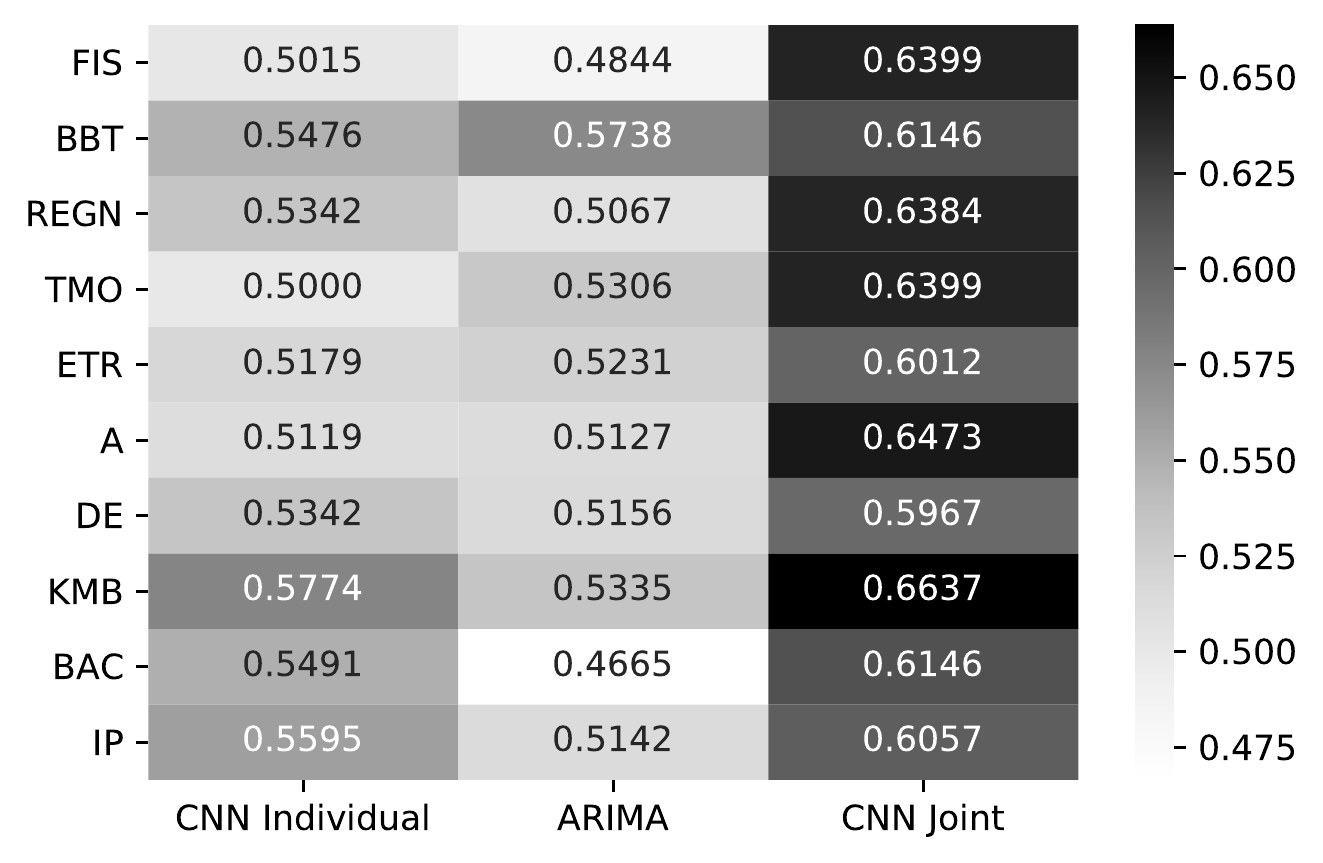}
\caption{Accuracy Scores}
\label{fig:accuracy_scores}
\end{figure}

The results of the forecast comparisons are available in Table \ref{table:metrics}. The metrics were averaged over the 10 stocks under study.\\
RMSE (root mean squared error) and SMAPE (symmetric mean absolute percentage error) are the reported regression metrics, while directional accuracy and F1 score describe the forecasts’ ability to get the directions right. We used both RMSE and SMAPE in order to show absolute and relative error measures as well. Accuracy and F1 score are commonly used metrics for binary classification.\\
The single-stock CNNs’ poor performance probably stems from the limited data volume. Neural networks excel when there’s a huge train set, and they might struggle with such data scarcity. Their unexploited complexity does more harm than good. For this reason, we have also compared our joint convolutional neural network to simple ARIMA models fitted to the individual volatility series.\\
Following \citet{hyndman2007automatic}, we used successive KPSS tests \citep{kwiatkowski1992testing} to choose the order of differencing. Then we used grid search to find the proper number of autoregressive and moving average terms between 0 and 3. We have chosen the best model based on AIC \citep{akaike1974new}.\\
Our CNN trained on multiple stocks outperformed ARIMA forecasts of the individual stock volatilities according to all metrics, even though the ARIMA parameters were chosen using a systematic procedure, while the CNN parameters were chosen rather arbitrarily. The convolutional neural network’s performance could probably have been further optimized by using grid search to find optimal hyperparameters.\\
The joint CNN model outperformed the single models in terms of forecast error and directional accuracy as well. Figure \ref{fig:rmse_scores} displays the average distance of forecasted and true values in terms of RMSE. Figure \ref{fig:accuracy_scores} shows directional accuracies.

\begin{table}[h!]
\centering
\begin{tabular}{ |c|c|c|c| }
 \hline
  & CNN Individual & ARIMA & CNN Joint \\
 \hline
 \multicolumn{4}{c}{Value Forecasts} \\
 \hline
 RMSE&0.0283&0.0261&0.0154 \\
 SMAPE&9.6324&4.9468&3.9358 \\
 \hline
 \multicolumn{4}{c}{Direction Forecasts} \\
 \hline
 Accuracy&0.5333&0.5161&0.6262 \\
 F1&0.5379&0.4182&0.6835 \\
 \hline
\end{tabular}
\caption{Evaluation metrics averaged over stocks}
\label{table:metrics}
\end{table}

\section{Conclusions and future perspectives}
We trained a one dimensional convolutional neural network on multiple stocks’ volatility rate history, and compared its forecasting performance to benchmark models trained on single series. We found that the deep learning method could take advantage of the multiplied data volume and produce better results, considering either value or direction forecasts and different measures of the goodness of the predictions. It suggests that the generality of stock prices allows a data expansion that might enable deep learning methods to outperform traditional time series models in (short-term) financial forecasting.\\
These findings open up research opportunities regarding the financial application of deep learning methods.\\
It should be explored if the results apply to different markets and to different forecasting horizons. For example, it would be worth examining if our jointly learned models can help forecasting volatilities of less frequently traded stocks. Or if it works with different data frequencies. We still used very small data---a few hundred stocks with daily price observations. Using intraday stock market data would seem more encouraging.\\
Also, further research should study, if similar joint machine learning models could be applied  to time series with higher diversity. Volatilities express a high degree of similarity, which justifies the model’s performance. However, learning general time series patterns might have a much broader scope.

\bibliography{bib}

\end{document}